# Aromatic Hexazine $[N_6]^{4-}$ Anion Revealed in the Complex Structure of the High-Pressure Potassium Nitride $K_9N_{56}$

Dominique Laniel[1,*], Florian Trybel[2], Yuqing Yin[1,3], Timofey Fedotenko[1], Saiana Khandarkhaeva[4], Andrey Aslandukov[1], Alexei I. Abrikosov[5], Talha Bin Masood[5], Carlotta Giacobbe[6], Eleanor Lawrence Bright[6], Konstantin Glazyrin[7], Michael Hanfland[6], Ingrid Hotz[5], Igor A. Abrikosov[2], Leonid Dubrovinsky[3], Natalia Dubrovinskaia[1,2].

**Affiliations:**

[1]Material Physics and Technology at Extreme Conditions, Laboratory of Crystallography, University of Bayreuth, 95440 Bayreuth, Germany

[2]Department of Physics, Chemistry and Biology (IFM), Linköping University, SE-581 83, Linköping, Sweden

[3]State Key Laboratory of Crystal Materials, Shandong University, Jinan 250100, China

[4]Bayerisches Geoinstitut, University of Bayreuth, 95440 Bayreuth, Germany

[5]Department of Science and Technology (ITN), Linköping University, SE-60174 Norrköping, Sweden

[6]European Synchrotron Radiation Facility, B.P.220, F-38043 Grenoble Cedex, France

[7]Photon Science, Deutsches Elektronen-Synchrotron, Notkestrasse 85, 22607 Hamburg, Germany

*Correspondence to dominique.laniel@uni-bayreuth.de.

**Abstract**

Recent high-pressure synthesis of pentazolates and subsequent stabilization of the aromatic $[N_5^-]$ anion at atmospheric pressure had an immense impact on nitrogen chemistry. Here, we present the first synthesis of an aromatic hexazine $[N_6]^{4-}$ anion realized in high-pressure potassium nitride $K_9N_{56}$ at 46 and 61 GPa. The extremely complex structure of $K_9N_{56}$ was solved based on synchrotron single-crystal X-ray diffraction and corroborated by density functional theory calculations. This result resolves a long-standing question of the aromatic hexazine stability and the possibility of its synthesis.

**Introduction**

The delocalization of electrons in aromatic systems leads to a significant enhancement of their structural stability [1–3] and enables the formation of key hydrocarbon ingredients of life [4–7]. In its simplest definition, an aromatic system is comprised of a cyclic, planar species with $4n+2$ $\pi$-electrons—criteria for aromaticity known as Hückel's rule [8]. This definition was later expanded to structural, magnetic, electronic, energetic, and reactivity indices [3] to account for aromaticity in systems with non-conventional and exotic geometries (*i.e.* Möbius aromaticity [9]). Although aromaticity was initially thought to be exclusive to carbon cycles, it has since been shown that numerous systems comprised of carbon heterocycles [10] and non-carbon cycles [11] can have an aromatic character.

Despite the surge of newly discovered homoatomic nitrogen species, including isolated integer and non-integer charged $[N_2]^{x-}$ dimers [12,13], the $[N_4]^{4-}$ tetranitrogen [14], the polymeric $[N_4]^{2-}_\infty$ [14–17], and N-frameworks [14,18–20], nitrogen aromaticity is thus far restricted to the $[N_5^-]$ pentazolate anion [21–23]. It is no coincidence that the synthesis of the $[N_5^-]$ species, as opposed to the others, has revolutionized our understanding of nitrogen chemistry in recent years: it has an increased stability due to its aromaticity and, therefore, an incredible potential for novel technological materials [24]. The application of high-pressure and high-temperature was decisive in the pentazolate-anion synthesis breakthrough, as its first bulk stabilization was achieved in $CsN_5$ by laser-heating of $CsN_3$ in $N_2$ at 60 GPa [22]. Although $CsN_5$ was found non-recoverable to ambient conditions, its synthesis was quickly followed by an ambient conditions approach to producing the pentazolate anion [25–27].

The six-member nitrogen ring, hexazine, if synthesized, would highly enrich nitrogen chemistry. In 1980, Vogler *et al.* [28] first reported the experimental synthesis of a neutral hexazine ring, or hexaazabenzene, through the photolysis of *cis*-$[Pt(N_3)_2(PPh_3)_2]$ at 77 K. However, this claim was later ruled improbable as theoretical calculations demonstrated that hexaazabenzene would decompose into three $N_2$ molecules due to the lack of an activation barrier [29]. Since then, a plethora of theoretical calculations has been performed to identify potential pathways to stabilise a hexazine ring as well as its most favorable geometry [30,31,40–44,32–39]. Although a variety of configurations were proposed, one that stands out from others is the aromatic hexazine anion $[N_6]^{4-}$ [30,41–43]. Up to this date, only the non-aromatic variant of the elusive $N_6$ ring has been reported, as an anion adopting an armchair configuration in $WN_6$ [45] and, in $K_2N_6$, as a planar anti-aromatic ($4n$ $\pi$-electrons) $[N_6]^{2-}$ species [46]. In both cases, however, the quality of the powder X-ray diffraction data was insufficient for a full structural refinement.

Here, we report the synthesis of the $K_9N_{56}$ solid at 46 and 61 GPa by laser heating $KN_3$ embedded into $N_2$. Its structure, determined using single-crystal X-ray diffraction, features the largest unit cell ever observed in solids produced above 1 GPa. Moreover, it contains the long-sought-after aromatic $[N_6]^{4-}$ species. Theoretical calculations reproduce the experimentally determined crystal structure of $K_9N_{56}$ and provide further insight into the properties of the compound and the aromatic character of the hexazine anion.

**Results and Discussion**

A screw-type BX90 diamond anvil cell (DAC) [47] was loaded with potassium azide ($KN_3$) and molecular nitrogen gas (~1200 bars), the latter serving as both a reagent and a pressure transmitting medium (see the Experimental Method section in the Supplementary Materials for the complete details). The sample was initially compressed to 46 GPa and laser-heated to a maximum temperature of 3200(200) K. After laser-heating, optical observations of the sample revealed changes in the sample's appearance; growing substantially in size (Figure S1). Raman spectroscopy measurements of the laser-heated regions of the sample no longer displayed the characteristic vibron of the azide anion around 1480 cm$^{-1}$, indicating a chemical transformation of the $KN_3$ compound (Figure S1), although no new sharp modes were observed. A first X-ray diffraction investigation of the laser-heated sample revealed diffraction lines that did not belong to any known compound (*i.e.* $KN_3$, $K_2N_6$, $K_3(N_2)_4$ [46]) or elemental phases ($\varepsilon$-$N_2$ [48], K-III [49]). At 46 GPa, a very large orthorhombic unit cell with lattice parameters of $a = 5.419(5)$ Å, $b = 34.73(3)$ Å and $c = 23.172(9)$ Å ($V = 4361(5)$ Å$^3$) was obtained from the data collected at the P02.2 beamline of PETRA III by performing single-crystal X-ray diffraction of the polycrystalline sample (SC-XRDp, see Supplementary Materials). However, the weak intensity of the reflections did not allow for a satisfactory structural analysis. The pressure was subsequently increased to 61 GPa and the sample laser-heated to a maximum temperature of 3400(200) K. The X-ray diffraction characterization of the reheated sample at both the ID11 and ID15 beamlines of the new EBS-ESRF—both beamlines having an extremely high flux—resulted in a significantly increased reflections' intensity. Using the SC-XRDp datasets collected at these beamlines, an orthorhombic unit cell with the lattice parameters of $a$ = 5.2380(6) Å, $b$ = 34.560(3) Å and $c$ = 23.2050(19) Å ($V$ = 4200.7(7) Å$^3$) at 61 GPa was determined and the space group assigned (*Ibam*, #72). As detailed in Table 1, containing part of the crystallographic data, a good $R_{int}$ value (6.39%) could be obtained upon data integration despite the large unit cell and weak scattering at high 2θ. Moreover, these X-ray diffraction data were sufficient for structural solution and refinement, unveiling a total number of 520 atoms in the unit cell, 72 K and 448 N, resulting in the $K_9N_{56}$ stoichiometry. According to the ICSD database, $K_9N_{56}$ has at 46 GPa the largest unit cell volume for any compound formed above 1 GPa.

**Table 1:** Crystallographic data for $K_9N_{56}$. Some parameters have both the experimental and the calculated value. The crystallographic data has been submitted under the deposition number CSD 2127463. The full crystallographic data can be found in Table S1.

|  | $K_9N_{56}$ |  |
| --- | --- | --- |
|  | **Exp.** | **Calc.** |
| Pressure (GPa) | 61 | 61 |
| Space group, # | *Ibam*, 72 | *Ibam*, 72 |
| Z | 8 | 8 |
| $a$ (Å) | 5.2380(6) | 5.499 |
| $b$ (Å) | 34.560(3) | 34.196 |
| $c$ (Å) | 23.2050(19) | 22.389 |
| $V$ (Å$^3$) | 4200.7(7) | 4210 |
| $R_{int}$; $R_1$; $wR_2$ (%) | 6.39; 7.47; 7.43 |  |

The structure model of $K_9N_{56}$ (Figure 1), derived from the SC-XRDp experiments, revealed that there are 448 nitrogen atoms in the unit cell, which form three kinds of structural groups: (i) planar $N_6$ rings, (ii) planar $N_5$ rings, and (iii) $N_2$ dimers. The four $N_6$ rings in the unit cell are crystallographically equivalent, whereas among the 56 $N_5$ rings and the 72 $N_2$ dimers there are four rings and seven dimers, correspondingly, which are crystallographically distinct. To help visualize this, non-equivalent rings are colored differently in Figure 1. A view along the *a*-axis (Figure 1 a) allows the best understanding of the complex crystal structure of $K_9N_{56}$. The positions of the centers of the planar $N_6$ rings (shown green) are fixed at the center of inversion (2/m). The two N1 atoms of the ring lie in the mirror (001) plane, and four N2 atoms are in the general position. The planar $N_6$ rings therefore have only one degree of freedom and are rotated about the *c*-axis at 45°. Their immediate chemical environment is shown in Figure 1 b-e. Crystallographically distinct $N_5$ rings are shown in four different colors in Figure 1 a. They build the major frame of the structure staking in infinite columns aligned along the *a*-axis (Figure 1 a). The columns either consist of rings of the same kind (those formed by N13, N14 and N15 atoms (red) and those formed by N16 through N20 atoms (orange)), or are comprised of alternating rings of different kinds (shown in pink and purple in Figure 1 a). The $N_2$ dimers (light blue), along with K atoms (light grey), fill in the space between the columns of $N_5$ rings. A single K5 atom occupies a general position. Other K atoms are in special positions with the symmetry *2* or *m*. Arrays of K atoms are all oriented along the *a*-axis.

The Kohn-Sham density functional theory (DFT)-based calculations reproduced the experimental atomic arrangement and the compound's dynamical stability (Figure S2) was demonstrated through *ab initio* molecular dynamics calculations (see Supplementary Materials for full details). A complete comparison between the DFT- and the SC-XRDp-derived crystal models, found to be in agreement, is provided in Table S1. The calculated electronic density of states revealed the metallic character of the compound (Figure S3).

According to the experimental model, $N_5$ rings have an average intramolecular bond length of 1.27(1) Å. The DFT-based structural relaxation leads to $N_5$ intramolecular bond lengths ranging from 1.28 to 1.31 Å (Figure S2), therefore in agreement with the experimental values considering their uncertainty. This is also in agreement with the bond lengths previously observed for $[N_5]^-$ rings at high pressure (between 1.26 and 1.32 Å in $NaN_5$ and $NaN_5 \cdot N_2$ at 47 GPa [23]). The majority of the $N_2$ dimers (shown blue) are close to being parallel to the *a*-axis and have an average intramolecular N-N distance of 0.98(3) Å based on the experimental data. This qualitatively agrees with the calculations, giving the values in the range between 1.10 and 1.12 Å, which is typical for a neutral nitrogen molecule [12,50,51]. Thus, assuming weak van der Waals interactions between $N_2$ molecules and the other constituents of $K_9N_{56}$, the latter can be considered as an insertion compound, akin to $NaN_5 \cdot N_2$ [23], $ReN_8 \cdot xN_2$ [16], $Hf_4N_{20} \cdot N_2$, $WN_8 \cdot N_2$ and $Os_5N_{28} \cdot 3N_2$ [18].

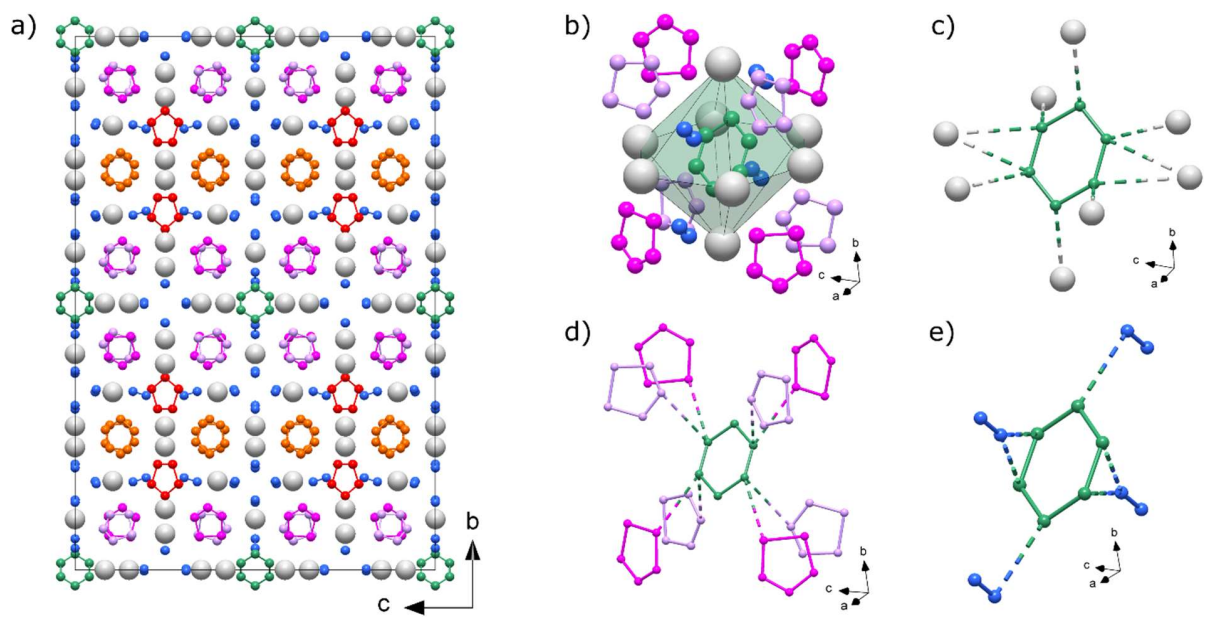

*Figure 1: Experimentally determined crystal structure of the $K_9N_{56}$ compound at 61 GPa. a) Unit cell viewed along the a-axis. b) Chemical environment of the $N_6$ ring, with an emphasis on the neighboring c) K atoms; d) $N_5$ rings; and e) $N_2$ dimers. The dashed lines represent the shortest intermolecular bonds between species, ranging from 2.266(17) to 2.629(24) Å in c), 2.542(18) to 2.649(19) Å in d) and 2.249(25) to 2.599(27) Å in e). The light grey spheres represent potassium atoms while the green and blue spheres are atoms forming, respectively, $N_6$ rings and $N_2$ dimers while the red, orange purple and pink represent N atoms producing $N_5$ rings. The crystallographic information for $K_9N_{56}$ is available for download on the CCDC database under deposition number CSD 2127463.*

The $N_6$ ring identified in $K_9N_{56}$ is the first hexazine unit detected based on an experimental structure refinement. According to the latter, there are four bonds with a length of 1.17(2) Å and two of 1.23(4) Å at 61 GPa. Within their uncertainty, these values are not discernible. According to the DFT model, the corresponding bond lengths are 1.32 and 1.28 Å—very similar to each other and intermediate between the values which are typical for the N-N single and double bonds [12]. They are also in agreement with the N-N bond lengths previously predicted for planar $N_6$ rings, for example, 1.37 Å in $Be_2N_6$ at 1 bar [40], 1.31 Å in $K_2N_6$ at 60 GPa [34], 1.31 Å in $K_2N_6$ at 50 GPa [52], 1.31 Å in $Li_2N_6$ at 50 GPa [35]. However, they differ from those predicted for armchair-shaped $N_6$ rings (*e.g.* 1.43 Å at 1 bar in $WN_6$ [53], 1.368 Å in $TeN_6$ at 90 GPa [36]) and for isolated, planar and cyclic $C_6$ anions (*e.g.* 1.50-1.51 Å in $Y_2C_3$ at 50 GPa [54]; 1.421 Å in $NaC_2$ at 100 GPa [55]).

The assignment of formal charges to each of the species in $K_9N_{56}$, whose formula can be written as $K_{72}[N_6]_4[N_5]_{56}[N_2]_{72}$, provides a crucial insight into the compound's crystal chemistry. As potassium is expected to have a formal charge of +1, 72 electrons should be split between the nitrogen units. The $N_2$ dimers have been suggested to be neutral based on their intramolecular bond length (see above). As the geometry of the $N_5$ rings in $K_9N_{56}$ is similar to that of currently known $N_5^-$ anions [23,25,26,56], then the 56 $N_5$ rings, each with a formal charge -1, account for 56 electrons. This leaves 16 electrons to be shared between the four crystallographically equivalent $N_6$ rings, implying a formal charge per ring of -4 and, thus, the assignment of charges in the chemical formula is as follows: $[K^+]_{72}[N_6^{4-}]_4[N_5^-]_{56}[N_2]_{72}$. The resonant structure of the $[N_6]^{4-}$ anions can be understood in the following way: each nitrogen atom in the ring has three *p*-electrons, of which two contribute to the bonding with two neighbours in the ring, and one contributes to the $\pi$-system. Thus, the $\pi$-system consists of these six $\pi$-electrons plus the four provided by the $K^+$ cations, resulting in $10\pi$-electrons. As such, the cyclic planar $[N_6]^{4-}$ hexazine anion, with $4n+2$ $\pi$ (*n*=2) delocalized electrons, effectively produces a $10\pi$-electron system and thus fulfills Hückel's rule for aromaticity [8]. The aromatic properties of the $N_6^{4-}$ hexazine anion are also supported by the following theoretical considerations. i) The electron density of the $[N_6]^{4-}$ ring (Figure 2) calculated in this work features an almost completely homogeneous charge distribution across the ring, implying a charge delocalization across it—a hallmark of aromaticity. ii) The geometry-based harmonic oscillator model of aromaticity (HOMA, see Supplementary Materials for detail) value [3], which can vary between 0 and 1 and where 1 corresponds to very strong aromaticity and 0 to very poor aromaticity, is equal to 0.96 for the hexazine anion in $K_9N_{56}$. That is comparable or even larger than the values of the known aromatic molecules pyrrole, selenophene and imidazole, having values of 0.86, 0.88 and 0.998, respectively [3]; and iii) previous theoretical calculations, including nucleus independent chemical shift (NICS) [33,41,43], also support aromaticity of the planar $[N_6]^{4-}$ cycle.

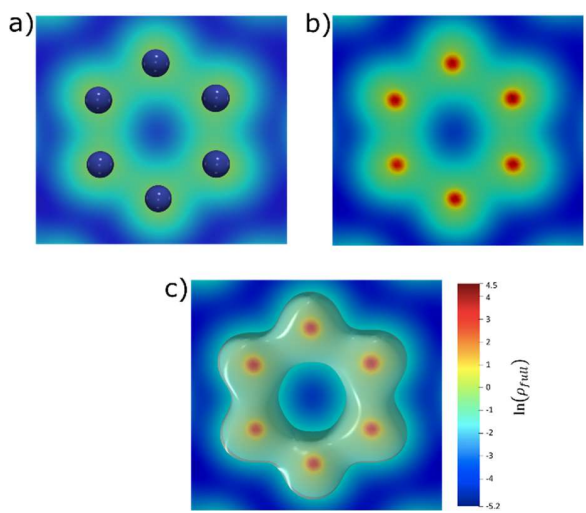

Figure 2: 2D slice through the all-electron charge density from the DFT-based calculations in the plane defined by the $N_6$ ring in $K_9N_{56}$, a) with and b) without the nitrogen atoms, drawn as blue spheres. c) The same $N_6$ ring as in b) but with an isoelectronic surface of 0.2 e$^-$/Å$^3$. The scale for the all-electron charge density is the same for all three figures and shown to the right of c). The almost completely homogenous nature of electronic density of the $N_6$ ring is further discussed in Figure S4.

The $K_9N_{56}$ compound, synthesized at 61 GPa, was decompressed to assess its stability domain. Powder XRD data from $K_9N_{56}$ could be collected down to 32 GPa (see Figure 3a), but single-crystal XRD data only to 41 GPa. Four pressure-volume points (two obtained on compression and the other two on decompression) are shown in Figure 3b. The measured unit cell volumes match well with theoretical calculations (see the dashed P-V curve in Figure 3 b), and the bulk modulus of $K_0 = 18(3)$ GPa ($K' = 4.8(3)$; $V_0 = 8178(50)$ Å$^3$) was determined on the basis of the *ab initio* calculations data. The bulk modulus of $K_9N_{56}$ is smaller than that of sodium pentazolate ($NaN_5$, $K_0 = 33.0(4)$ GPa [23]), as expected from the presence of the van der Waals-bonded $N_2$ dimers. Below 32 GPa, solely the diffraction lines of $\varepsilon$-$N_2$ could be observed, suggesting the decomposition or amorphization of $K_9N_{56}$. Again, considering the many neutral and loosely bonded $N_2$ dimers, the decomposition of $K_9N_{56}$ was expected.

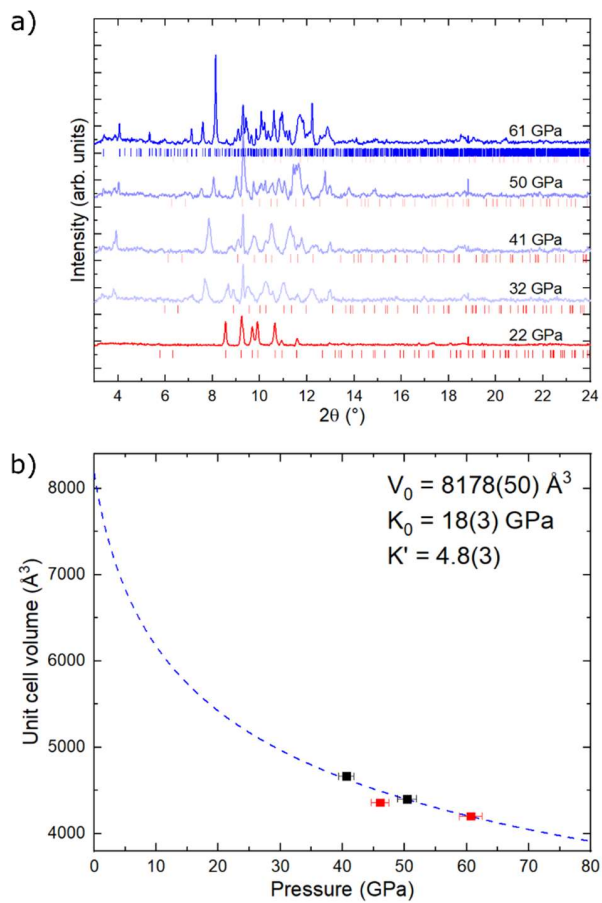

Figure 3: a) Diffraction patterns of $K_9N_{56}$ collected during its decompression. At 22 GPa, only the diffraction lines of $\varepsilon$-$N_2$ are observed. The blue and red tick lines correspond to the position of the diffraction lines of $K_9N_{56}$ and $\varepsilon$-$N_2$, respectively. For clarity of the figure, the tick lines for $K_9N_{56}$ are only shown under the upper diffraction pattern (61 GPa). b) The pressure dependence of the unit cell volume of $K_9N_{56}$. The red and black datapoints were collected under compression and decompression, respectively. The dashed blue line was obtained from theoretical calculations used for determining the values of the parameters of a 3$^{rd}$ order Birch-Murnaghan equation of state The volume and unit cell parameters obtained from the experimental pressure points are shown in Table S2.

**Conclusions**

We synthesized a $K_9N_{56}$ compound by the laser-heating of $KN_3$ embedded in molecular nitrogen at pressures of 46 and 61 GPa. The unit cell of this compound with 520 atoms and a volume of 4361(5) Å$^3$ at 46 GPa is the largest among the solids synthesized above 1 GPa, whose structure was solved and refined. Single-crystal synchrotron X-ray diffraction revealed that the structure of $K_9N_{56}$ is composed of a complex arrangement of $[N_6]^{4-}$ and $[N_5]^{-}$ rings as well as neutral nitrogen dimers. According to Hückel's rule, the $[N_6]^{4-}$ ring is aromatic as it is cyclic, planar and has $4n+2$ $\pi$-electrons ($10\pi$-system). Its aromaticity is further supported by bond lengths considerations and by calculations of the electronic charge density. The computed electron density of states of $K_9N_{56}$ shows a metallic character of the material with a low electron density at the Fermi energy that agrees well with the lack of distinctly measurable Raman modes. Under pressure decrease, $K_9N_{56}$ was observed down to 32 GPa, after that its diffraction lines are no longer detectable.

These results emphasize that single-crystal X-ray diffraction on micro samples (SC-XRDp) obtained using the extreme-brilliance radiation of the fourth-generation synchrotrons is crucial for solving very complex structures of solids at high pressures. This study also provides a striking example against the trope of structural simplicity at high densities. Resulting in the synthesis of the long-sought-after aromatic hexazine anion, the second known aromatic homoatomic nitrogen species after the $[N_5]^{-}$ pentazole anion, this work can stimulate further exploration of nitrogen chemistry in the search of novel nitrogen-based technological materials.

**Methods**

*A. Experimental Method*

A BX90-type screw-driven diamond anvil cell (DAC) [47] was equipped with 250 μm culet diamond anvils. A rhenium gasket with an initial thickness of 200 μm was indented down to ~25 μm and a sample cavity of ~120 μm in diameter was laser-drilled at the center of the indentation. The DAC was loaded with potassium azide, $KN_3$, as well as molecular nitrogen gas (~1200 bars), acting as a reagent and a pressure transmitting medium. The *in-situ* pressure was measured using the first-order Raman mode of the stressed diamond anvils [57]. Double-sided sample laser-heating was performed at our home laboratory at the Bayerisches Geoinstitut [58] with $KN_3$ employed as the laser absorber. Temperatures were measured with an accuracy of ±200 K, using the thermoemission produced by the laser-heated samples [58].

Synchrotron X-ray diffraction measurements of the compressed samples were performed at ID11 ($\lambda$ = 0.2852 Å) and ID15 ($\lambda$ = 0.41015 Å) of the EBS-ESRF as well as the P02.2 beamline ($\lambda$ = 0.2895 Å) of PETRA III. In order to determine the position of the polycrystalline sample on which the single-crystal X-ray diffraction (SC-XRDp, "p" standing for "polycrystalline samples") acquisition is obtained, a full X-ray diffraction mapping of the pressure chamber was achieved. The sample position displaying the most and the strongest single-crystal reflections belonging to the phase of interest was chosen for the collection of single-crystal data, collected in step-scans of 0.5° from −38° to +38°. The CrysAlisPro software [59] was utilized for the single crystal data analysis. The analysis procedure includes the peak search, the removal of the diamond anvils' and other 'parasitic' signal contributions, finding reflections belonging to a unique single crystal, the unit cell determination, and the data integration. The crystal structures were then solved and refined using the OLEX2 and JANA2006 software [60]. The SC-XRDp data acquisition and analysis procedure was previously demonstrated and successfully employed [17,61,62]. The full details of the method can be found elsewhere [63]. Powder X-ray diffraction (pXRD) measurements were also performed to verify the sample's chemical homogeneity and identify the presence of the $K_{72}N_{448}$ phase when sufficient quality single-crystal data could no longer be obtained. The powder X-ray data was integrated with the Dioptas software [64].

Confocal Raman spectroscopy measurements were performed with a LabRam spectrometer equipped with a ×50 Olympus objective. Sample excitation was accomplished using a continuous He-Ne laser (632.8 nm line) with a focused laser spot of about 2 μm in diameter. The Stokes Raman signal was collected in a backscattering geometry by a CCD coupled to an 1800 l/mm grating, allowing a spectral resolution of approximately 2 cm$^{-1}$. A laser power of about 4.6 mW incident on the DAC was employed.

### B. *Density functional theory calculations*

Kohn-Sham density function theory based structural relaxations and electronic structure calculations were performed with the Quantum Espresso package [65–67] using the projector augmented wave approach [68]. We used the generalized gradient approximation by Perdew-Burke-Ernzerhof (PBE) [69] for exchange and correlation, with the corresponding potential files: for K the 2p electrons and lower and for N the 1s electrons are treated as scalar-relativistic core states. Convergence tests with a threshold of 1 meV per atom in energy and 1 meV/Å per atom for forces led to a Monkhorst-Pack [70] k-point grid of 4x4x4 for structural relaxations and 8x8x8 for the calculations of electronic properties in the primitive unit cells. We used the cutoff energy of 100 Ry for the expansion of the wave function. Molecular dynamics (AIMD) simulations were performed with the Vienna *ab initio* simulation package (VASP) [71–73], employing a 2x2x2 Monkhorst-Pack k-point grid and a cutoff energy of 450 eV with PBE potentials, where for N and K, respectively, the 1s electrons and the 3s electrons as well as those in lower energy orbitals, were treated as core states.

We performed variable cell relaxations (lattice parameters and atomic positions) on the experimental structure of $K_9N_{56}$ to optimize the atomic coordinates and the cell vectors until the total forces were smaller than $10^{-4}$ eV/Å per atom and the deviation from the experimental pressure was below 0.1 GPa. Using an initial electronic smearing, which can be interpret as an electronic temperature ($T_{el}$), of 0.005 Ry (~800 K) in our static calculations leads to a difference of $\Delta d \approx 10\%$ between the six intramolecular bond lengths of the $N_6$ ring. This is significantly larger than the difference of 4.87% found in the experimental structure refinement. The experimental unit cell volume was reproduced with a deviation of ~0.2%; for the lattice constants we obtained deviations of +4.5%, -1.3% and -3.3% for *a*, *b* and *c*, respectively. The calculated atomic coordinates do not perfectly match a *Ibam* (#72) symmetry, that was derived from the experimental data, but are nonetheless quite close to fitting it. Calculating the electronic density of states (eDOS) shows that the compound is metallic (Figure S3). By increasing the $T_{el}$ stepwise from 0.005 Ry (~800 K) to 0.05 Ry (~8000 K) in our static calculations to get an approximation on the finite temperature behavior of the system [74], it was found that the structure fully adopts the *Ibam* space group for $T_{el} \geq 4000$ K (0.025 Ry). With increasing $T_{el}$, the eDOS undergoes no significant modification aside from becoming smoother (Figure S3). The interatomic distances in the planar $N_6$ rings become nearly equal with larger $T_{el}$ values (*e.g.* $\Delta d_{6000K} \approx 4\%$, $\Delta d_{8000K} \approx 2.5\%$). This can be interpreted [74] as an indication for a thermal stabilization of a high symmetry structure containing planar $N_6$ rings with six equal interatomic distances.

Due to the extremely large size of the unit cell, it was not feasible to calculate a fully converged phonon band structure for $K_9N_{56}$. Instead, we opted to perform an *ab initio* molecular dynamics (AIMD) simulation at the experimental temperature of 300 K to obtain further insight on the effect of lattice vibrations on the symmetry of the $N_6$ rings and the stability of the compound [75]. We ran 5600 steps of the AIMD with a time step of 1 fs and an electronic smearing of 0.1 eV (1200 K). We treated the first 800 steps as equilibration steps and therefore did not take these steps into account in the analysis. The modification of the structure as a function of the simulation time was monitored by calculating the average interatomic distances for the $N_6$ rings, the $N_5$ rings and the $N_2$ dimers in each time step (Figure S2a). The bond lengths undergo fluctuations in the order of ±0.025 Å, but no significant structural modifications or movement of atoms was observed. However, the $N_6$ rings in the averaged structure show a very small deviation from being completely planar. Analyzing individual AIMD steps suggests that the rings show a bending mode, which is not fully averaged out in the finite number of steps. Most importantly, the rings do not adapt an armchair configuration in any of the analyzed AIMD steps. In the AIMD calculations, the components of the stress tensor do not diverge and off-diagonal elements remain on average zero (Figure S2b). We found a slightly higher fluctuation in $\sigma_{xx}$ (in cartesian coordinates), which is most likely a consequence of the *a* lattice constant (~5.2 Å) being significantly smaller than the *b* and *c*, both > 20 Å (Table S1). Calculating the average structure, we found interatomic distances in the $N_6$ rings to have a $\Delta d_{AIMD} \approx 3.5\%$, relatively close to the value of 4.87% found in the experimental structure refinement. Comparison of the average interatomic distances in the rings and the dimers between the AIMD and the relaxations, we find that a static calculation with $T_{el}$ of 6000 K best matches the average AIMD distances of the $N_6$ and $N_5$ rings. This can be viewed as a confirmation that static calculations with an increased $T_{el}$ capture some effects of the lattice dynamics [74]. The interatomic distances are in general mildly larger in the AIMD, which is most likely caused by the ionic motions. In general, all relaxations with $T_{el} \geq 4000$ K led to similar structures and are in satisfactory agreement with the average structures from the AIMD as well as with the experimental refinement. As such, all further calculations were performed with the $T_{el} = 6000$ K structure.

We used the geometry-based aromaticity index HOMA (Harmonic oscillator model of aromaticity), described by Krygowski *et al.* [3], to obtain a measure of the aromatic character of the $N_6$ rings in $K_9N_{56}$. The HOMA is defined as:

$$HOMA = 1 - \alpha(R_{opt} - R_{av})^2 - \frac{1}{n}\sum_i \alpha(R_{av} - R_i)^2,$$

where $R_{opt}$ denotes the length of the bond for which extension to the single bond and compression to the double bond costs energetically the same, $R_{av}$ is the average bond length in the $N_6$ ring, $\alpha$ is a normalization constant depending on the elements involved in he bonds and $R_i$ is the bond length of the $i^{th}$ bond. For the $N_6$ ring in $K_9N_{56}$, according to the DFT-relaxed model, we find an $R_{av}$ = 1.294 Å. We used values $R_{opt}$ = 1.309 Å and α = 130.33 from Ref. [3], which were determined for typical N-N and N=N bonds at ambient conditions. While these values were obtained at ambient conditions, they are not expected to change significantly due to the very high incompressibility of homoatomic nitrogen species, as previously demonstrated in Ref. [12]. We obtained a HOMA value of 0.96 which is close to the ideal value of 1 and therefore a strong indicator for the aromatic character of the $N_6$ rings.

Furthermore, we used the relaxed structure to calculate an equation of state (EOS) by isotropically compressing the simulation cell and relaxing the atomic positions until the remaining forces are $< 5 \cdot 10^{-3}$ eV/Å. A third order Birch Murnagham EOS was fitted to the calculated energies versus volume points and obtain $K_0$ = 18(3) GPa, $K'$ = 4.8(3) and $V_0$ = 8178(50) Å$^3$ (Figure 3). The EOS is in good agreement with experimental data obtained during compression and decompression.

**Acknowledgments**

The authors acknowledge the Deutsches Elektronen-Synchrotron (DESY, PETRA III) and the European Synchrotron Radiation Facility (ESRF) for provision of beamtime at the P02.2 and, ID15b and ID11 beamlines, respectively. D.L. thanks the Alexander von Humboldt Foundation and the Deutsche Forschungsgemeinschaft (DFG, project LA-4916/1-1) for financial support. N.D. and L.D. thank the Federal Ministry of Education and Research, Germany (BMBF, grant no. 05K19WC1) and the Deutsche Forschungsgemeinschaft (DFG projects DU 954–11/1, DU 393–9/2, and DU 393–13/1) for financial support. Support from the Swedish Research Council (VR) Grant No. 2019-05600, the Swedish Government Strategic Research Areas in Materials Science on Functional Materials at Linköping University (Faculty Grant SFO-Mat-LiU No. 2009 00971) and SeRC, and the Knut and Alice Wallenberg Foundation (Wallenberg Scholar Grant No. KAW-2018.0194) is gratefully acknowledged. Computations were performed on resources provided by the Swedish National Infrastructure for Computing (SNIC) at the PDC Centre for High Performance Computing (PDC- HPC) and the National Supercomputer Center (NSC).